\documentstyle[12pt,epsf]{article}

\baselineskip=\normalbaselineskip

\ifx\TwoupWrites\UnDeFiNeD\else\target{\magstepminus1}{11.3in}{8.27in}
\source{\magstep0}{7.5in}{11.69in}\fi
\newfont{\fourteencp}{cmcsc10 scaled\magstep2}
\newfont{\titlefont}{cmbx10 scaled\magstep3}
\newfont{\authorfont}{cmcsc10 scaled\magstep1}
\newfont{\fourteenmib}{cmmib10 scaled\magstep2}
\skewchar\fourteenmib='177
\newfont{\elevenmib}{cmmib10 scaled\magstephalf}
\skewchar\elevenmib='177
\makeatletter
\newcommand\nonsequentialeqnum{
\@addtoreset{equation}{section}
\def\theequation{\arabic{section}.\arabic{equation}}}
\newif\ifp@bblock \p@bblocktrue
\newcommand\nopubblock{\p@bblockfalse}
\newcommand\topspace{\hrule height 0pt depth 0pt \vskip}
\newcommand\p@bblock{\begingroup \tabskip=\hsize minus \hsize
\baselineskip=1.5\ht\strutbox \topspace-2\baselineskip
\halign to\hsize{\strut ##\hfil\tabskip=0pt\crcr
\the\Pubnum\crcr\the\date\crcr}\endgroup}
\renewcommand\titlepage{\ifx\TwoupWrites\UnDeFiNeD\null
\vspace{-1.7cm}\fi
\vskip0.6cm
\ifp@bblock\p@bblock \else\hrule height 0pt \relax \fi}
\makeatother
\newtoks\date
\newtoks\Pubnum
\newtoks\pubnum
\Pubnum={
~ \crcr
~ \crcr
YITP-99-9 \crcr
RUP-99-2 \crcr
hep-th/9902169 \crcr
{}February 1999
}
\newcommand{\frontpageskip}{\vspace{12pt plus .5fil minus 2pt}}
\renewcommand{\title}[1]{\frontpageskip
\begin{center}{\titlefont #1}\end{center}\par}
\renewcommand{\author}[1]{\frontpageskip\par\begin{center}
{\authorfont #1}\end{center}
\nobreak
}

\newcommand{\address}[1]{\par\begin{center}{\sl #1}\end{center}\par}

\renewcommand{\thanks}[1]{\footnote{#1}}
\renewcommand{\abstract}{\par\frontpageskip\centerline{
\fourteencp Abstract}
\vspace{8pt plus 3pt minus 3pt}}
\begin{document}

\titlepage

%\vskip 3cm

\renewcommand{\thefootnote}{\fnsymbol{footnote}}
\title{
\protect\Large{\protect\bf Comments on D-Instantons in $c<1$
Strings}
}

\author{
Masafumi Fukuma${}^{1\,}$\thanks{
e-mail address: {\tt fukuma@yukawa.kyoto-u.ac.jp}}
and~
Shigeaki Yahikozawa${}^{2\,}$\thanks{
e-mail address: {\tt yahiko@rikkyo.ac.jp}}
}

\address{
${}^1$
Yukawa Institute for Theoretical Physics\\
Kyoto University, Kyoto 606-8502, Japan \\
~\\
${}^2$
Department of Physics\\
Rikkyo University, Tokyo 171-8501, Japan \\
}

\newcommand{\bc}{\begin{center}}
\newcommand{\ec}{\end{center}}
\newcommand{\bl}{\begin{flushleft}}
\newcommand{\el}{\end{flushleft}}
\newcommand{\bi}{\begin{itemize}\begin{enumerate}}
\newcommand{\ei}{\end{enumerate}\end{itemize}}
\newcommand{\bt}{\begin{tabbing}}
\newcommand{\et}{\end{tabbing}}
\newcommand{\np}{\newpage}
\newcommand{\pset}{\setcounter{page}{1}}
\newcommand{\wh}[1]{w^{({#1})}}

\newcommand{\lettersection}[1]{\noindent
\underline{\bf{{#1}}}
}

\renewcommand{\thefootnote}{\arabic{footnote}}
\setcounter{footnote}{0}
\newcommand{\cleqn}{\setcounter{equation}{0} \indent}
\renewcommand{\theequation}{\arabic{equation}}
\newcommand{\beqa}{\begin{eqnarray}}
\newcommand{\eeqa}{\end{eqnarray}}
\newcommand{\n}{\nonumber}
\newcommand{\nn}{\nonumber \\ }
\newcommand{\eq}[1]{(\ref{#1})}
\newcommand{\bC}{{\bf C}}
\newcommand{\cD}{{\cal D}}
\newcommand{\cH}{{\cal H}}
\newcommand{\cO}{{\cal O}}
\newcommand{\Psid}{\Psi^\dagger}
\newcommand{\norm}[1]{{\parallel {#1} \parallel}^2}
\newcommand{\nnorm}[1]{{{\parallel {#1} \parallel}^{\prime\,2}_l}}
\newcommand{\del}{\partial}
\newcommand{\db}{{\bar{\delta}}}
\newcommand{\gbar}{{\bar{g}}}
\newcommand{\dl}
{\left[\,\frac{dl}{\,l\,}\,\right]}
\newcommand{\Det}{\,\mbox{Det}\,}
\newcommand{\Tr}{\,\mbox{Tr}\,}
\newcommand{\ldot}{\dot{l}}
\newcommand{\const}{{\rm const.\ }}
\newcommand{\bra}[1]{\left\langle\,{#1}\,\right|}
\newcommand{\ket}[1]{\left|\,{#1}\,\right\rangle}
\newcommand{\bracket}[2]{
\left\langle\left.\,{#1}\,\right|\,{#2}\,\right\rangle}
\newcommand{\vev}[1]{\left\langle\,{#1}\,\right\rangle}
\newcommand{\bZ}{{\rm {\bf Z}}}
\newcommand{\svac}{\bra{\sigma}}
\newcommand{\pvac}{\ket{\Phi}}
\newcommand{\rphi}{\varphi}
\newcommand{\rphih}{\hat{\rphi}}
\newcommand{\phih}{\hat{\phi}}
\newcommand{\dphi}{\del\rphi}
\newcommand{\dphih}{\del\hat{\phi}}
\newcommand{\cb}{\bar{c}}
\newcommand{\gint}{\oint^{\,p}}
\newcommand{\dds}{\frac{d s}{2\pi i}}
\newcommand{\ddz}{\frac{d\zeta}{2\pi i}}
\newcommand{\gst}{g_{\rm st}}
\newcommand{\gh}{\hat{g}}
\newcommand{\tr}{{\rm tr}}

%%%%%%%%%%%%%%%%%%%%%%%%%%%%%%%%%%%%%%%%%%%%%%%
% Abstract
%%%%%%%%%%%%%%%%%%%%%%%%%%%%%%%%%%%%%%%%%%%%%%
\begin{abstract}

We suggest that the boundary cosmological constant $\zeta$ in $c<1$
unitary string theory be regarded as the one-dimensional
complex coordinate of the target space
on which the boundaries of world-sheets can live.
{}From this viewpoint we explicitly construct 
analogues of D-instantons which satisfy 
Polchinski's ``combinatorics of boundaries.''
We further show that our operator formalism developed in the
preceding articles is powerful in evaluating D-instanton effects,
and also demonstrate for simple cases that these effects
exactly coincide with the stringy nonperturbative effects found
in the exact solutions of string equations.

\bl
%{\bf PACS no.\ }: {\tt 11.25.Pm, 11.25.Sq}\\
%{\bf Key Words}: {\tt Noncritical string theory, %Nonperturbative
%techniques, Soliton, \\
%Schwinger-Dyson equation, D-Instanton}
\el

\end{abstract}

%%%%%%%%%%%%%%%%%%%%%%%%%%%%%%%%%%%%%%%%%%%%%%%
% Main Part
%%%%%%%%%%%%%%%%%%%%%%%%%%%%%%%%%%%%%%%%%%%%%%%

Recent progress has revealed the vital role of D-branes \cite{p}
in the nonperturbative aspects of string theory.
{}For understanding nonperturbative dynamics which arises from the
D-branes,
it has been particularly important to consider the combinatorics
of boundaries
of world-sheets.
In fact,
Polchinski made an interesting observation
sometime ago \cite{p-g}
that the stringy nonperturbative effects
of the form $e^{-C\,/g}$ \cite{s}
($g$ is the coupling constant of closed strings)
could be explained in terms of the combinatorics of boundaries
in a target space with D-instanton background.
The point is that
when summing up connected diagrams in such target space,
one should take into account not only connected world-sheets,
but also configurations where disconnected world-sheets are
attached together at a D-instanton.
Then in the weak coupling limit, the dominant contribution comes
from the disk amplitude $-C/g$ for each world-sheet,
which will be accumulated to the form $e^{-C\,/g}$ 
after summing over the number of attached world-sheets.

In order to investigate such ``many-boundary systems," however,
it should be more natural and convenient to use quantum string fields
which can create and annihilate the boundaries of world-sheets.
{}Furthermore, if we can introduce the quantum fields
that create loop boundaries of Dirichlet type,
then the interactions and combinatorics
of D-branes can be dealt with in a second-quantized way.
Though for critical strings it seems difficult
to fully carry out this program in string-field language 
at hand,\footnote{
An attempt in critical string field theory can be found in \cite{hh},
where the interaction between a D-brane and a closed string field
are investigated by coupling the string field to the boundary state.
}
there might be a chance to do it for $c\leq 1$ unitary case,
because almost all the dynamical variables can be gauged away
with 2D diffeomorphism,
leaving the external lines only zero-modes (the positions).
This implies that the string field theory could reduce to
a local field theory in this case.
We will show that actually the string field theory for
$c=1-6/p(p+1)$ $(p=2,3,...)$
reduces to a chiral 2D conformal field theory
by identifying the target-space coordinate with
the so-called boundary cosmological constant $\zeta$.
In particular, the solitons $D_{ab}$ constructed
by the present authors \cite{fy1,fy2} (to be described below)
are interpreted as analogues of D-instantons and shown to satisfy
combinatorics similar to that of Polchinski.

We start our discussion with recalling that noncritical $c=1$ string
theory is a $D=2$ critical string theory with linear dilaton
background \cite{m}:
$G_{\mu \nu}(X)=\eta_{\mu \nu}={\rm diag}(-1,+1)$,
$ B_{\mu \nu}(X)=0$ and
$\Phi(X)=(Q/2) X^{1}$,
where $X^\mu ~(\mu =0,1)$ are coordinates of target space, and
$G_{\mu \nu}(X)$, $B_{\mu \nu}(X)$ and $\Phi(X)$ are, respectively,
the background metric, antisymmetric 2-tensor and dilaton.
Upon requiring the NL$\sigma$ model with this background
to be invariant under the Weyl transformation
as well as the 2D diffeomorphism,
the value of $Q$ is uniquely determined up to sign
(which can be absorbed into a redefinition of $X^1$) as
$Q=\sqrt{2(26-D)/3\alpha'}=4/\sqrt{\alpha'}$,
and this solution has no higher $\alpha'$ corrections
since the system is essentially Gaussian
(Hereafter we take the CFT unit, $\alpha'=2$).
The NL$\sigma$-model action with this linear dilaton term has
exactly the same form with that of Liouville gravity coupled to
$c=1$ conformal matter
if a spatial coordinate is regarded as Liouville field, $\phi=X^{1}$:
\beqa
S[X^\mu ]\,=\,\int d^2\sigma \sqrt{\gh}
\left\{\frac{1}{8\pi}\gh^{ab}
\eta_{\mu \nu}\,\del_a X^\mu \del_b X^\nu
\,+\,\frac{\hat{R}}{8\pi}\,Q^\mu X_\mu 
\right\},
\eeqa
with $Q^\mu =(0,Q)$.
(Here we have dropped the boundary term, or taken the world-sheet
metric such that its geodesic curvature vanishes.)
In fact, this is obtained \cite{ct-gn,kpz-ddk}
if we impose only the 2D diffeomorphism on a NL$\sigma$ model
with a scalar field $X^0$ (with negative metric)
and express the world-sheet metric as $g_{ab}=e^\phi \gh_{ab}$
which fluctuates around a fixed metric $\gh_{ab}$.
{}Furthermore, the $c<1$ case can be realized from $c=1$ ($D=2$)
by making a Lorentz boost on $(X^0,X^1)\rightarrow
(X^0\cosh\omega-X^1\sinh\omega,\,-X^0\sinh\omega+X^1\cosh
\omega)$
as well as on $Q^\mu $
with hyperbolic angle $e^\omega=(\sqrt{1-c}+\sqrt{25-c})/2\sqrt{6}$
\cite{lz},
and the corresponding NL$\sigma$-model action with vanishing
cosmological constant would be (after the Wick rotation $X=iX^0$):
\beqa
S[\phi,X]\,=\,\int d^2\sigma \sqrt{\gh}
\left\{\frac{1}{8\pi}\gh^{ab}
\left(\del_a \phi \,\del_b \phi + \del_a X\,\del_b X\right)
\,+\,\frac{\hat{R}}{4\pi}
\left(\frac{Q}{2}\phi\,+\,i \alpha_0 X \right)
\right\},
\eeqa
where $c=1-12\alpha_0^2$ and now $Q=\sqrt{(25-c)/3}$.
Note that after the Wick rotation
it has an invariance under the shift
$X\rightarrow X+2\pi/\alpha_0$
(including the case when boundaries exist).

In 2D gravity we can introduce the macroscopic-loop operator
$\tilde{\Psi}(l)$ \cite{bdss} that is defined effectively
as creating a loop boundary of length $l$
on the random surface.
Its Laplace transform is defined with the boundary cosmological
constant
$\zeta$ as
$\Psi(\zeta)~\equiv~\int_0^\infty dl\,e^{-l\zeta}\,\tilde{\Psi}(l)$,
and when $c=1-6/p(p+1)$ it is expanded around $\zeta=\infty$
(equivalently $l=0$) as
\beqa
\Psi(\zeta) ~=~\frac{1}{p}\,\sum_n \cO_n \zeta^{-n/p-1}, \label{3}
\eeqa
where $\cO_n$'s are scaling operators
and have the following form \cite{ct-gn,kpz-ddk}:
\beqa
\cO_n~=~\int d^2\sigma \sqrt{\gh}\,\,
e^{\beta_n \phi(\sigma)}\,\,e^{i\alpha_n X(\sigma)}, \label{4}
\eeqa
with $\beta_n=-\alpha_0(n-2p-1)$
and $\alpha_n=-\alpha_0(n-1)$ \cite{kpz-ddk}.
Note that these operators are also invariant under the shift
$X\rightarrow X+2\pi/\alpha_0$.
This implies that the external on-shell states effectively live on
a target space with the $X$-direction compactified with
radius $1/\alpha_0$,
while the bulk of the world-sheet lives whole in the 2D target space.
This is a specialty of the discrete unitary series.

Now we are going to argue that this $\zeta$ could be regarded
as a coordinate of target space.
One justification comes from the 2D-gravity side.
In fact, in the matrix-model regularization of this system,
the $\Psi(\zeta)$ is obtained as the continuum limit of
the resolvent of a random matrix $M$,
$f(P)=\tr\,(P-M)^{-1}$,
by setting $P=P_c\,e^{\zeta a}$ with $P_c$ the convergence radius
of $\vev{f(P)}$ around $P=\infty$ and $a$ the lattice spacing.
On the other hand, according to our experience in any matrix models
which generate random surfaces in $c$-dimensional target space
(see \cite{kazakov} for example),
the argument of the resolvent is always identified with an extra
coordinate of target space (thus totally $D=c+1$-dimensions).
This motivates us to identify $\zeta$ with a coordinate of target
space.

This idea can be elaborated more if we stand on the string-theory side.
In fact, substituting eq.\ \eq{4} into \eq{3},
we have
\beqa
\Psi(\zeta)&\sim&\int d^2\sigma \sqrt{\gh}\,\sum_n\,
e^{-n\alpha_0\phi(\sigma)}\,e^{-i n\alpha_0 X(\sigma)}\,
\zeta^{-n/p}\nn
&\sim& \int d^2\sigma \sqrt{\gh}\,\,
\frac{1}{\zeta^{1/p}\,-\,e^{-\alpha_0 (\phi(\sigma)+i X(\sigma))}}.
\label{5}
\eeqa
Here we have used
$e^{-n\alpha_0 (\phi(\sigma)+i X(\sigma))}\,
e^{-m\alpha_0 (\phi(\sigma)+i X(\sigma))}
\sim e^{-(n+m)\alpha_0 (\phi(\sigma)+i X(\sigma))}$,
since the singularities from contact interaction
both for $\phi$ and $X$ cancel each other.
We also neglected possible coefficients on the right hand side
of \eq{4} which would not give essential changes
to the singularity behavior in \eq{5}.
Noticing that the whole expression includes
only $\zeta$ (not $\bar{\zeta}$),\footnote{
This is also a specialty of $c<1$.
{}For $c=1$ there will appear both $\zeta$ and $\bar{\zeta}$
as well as extra discrete states.
The relationship between $c=1$ and $c<1$ is reminiscent
of the one between AdS$_3$ and AdS$_2$ in the AdS/CFT
correspondence
\cite{gksh}.
}
and also that the integrand is a delta function in a complex plane,
we thus see that the role of $\Psi(\zeta)$ is substantially
to create a loop boundary in one-dimensional complex target space
with the (exponentiated) coordinate $\zeta$.
{}Furthermore, due to the form $\zeta^{1/p}$, any observables
should be invariant under the rotation
$\zeta \rightarrow e^{2\pi ip}\zeta$ which corresponds to the shift
$X \rightarrow X-2\pi/\alpha_0$.
Although the above arguments might require to be refined more,
we may conclude that the on-shell boundaries
live on a complex one-dimensional target space
with a chiral coordinate $\zeta$ with periodicity
$\zeta\rightarrow e^{2\pi ip}\zeta$,
and the role of the operator $\Psi(\zeta)$
is to pin the world-sheet at the point
$\zeta^{1/p}=e^{-\alpha_0 ( \phi+i X)}.$

An operator formalism of $c<1$ string field theory
is developed in \cite{fy1,fy2}.
There the string field $\Psi(\zeta)$ is expressed as the first
derivative of
a scalar field $\rphi_0(\zeta)$, and has the following mode expansion
under the $\bZ_p$-twisted vacuum $\ket{\hat{\sigma}}$:
$\Psi(\zeta)\equiv\dphi_0(\zeta)=(1/p)\sum_{n\in\bZ}
\alpha_n \zeta^{-n/p-1}$
with $[\alpha_n,\,\alpha_m]=n\,\delta_{n+m,0}$
and $\alpha_n\ket{\hat{\sigma}}=0$ $(n\geq 0)$.
It thus raises a monodromy among $p$ scalar fields $\dphi_a(\zeta)$
$(a=0,1,...,p-1)$ as $\dphi_a(e^{2\pi i}\zeta)=\dphi_{[a+1]}(\zeta)$
with $[a]\equiv a$ (mod $p$).
The correlation functions (generally disconnected) are then given by
\beqa
\vev{\Psi(\zeta_1)\cdots\Psi(\zeta_n)}
&\equiv&\int_0^\infty
dl_1\cdots dl_n\,e^{-l_1\zeta_1-\cdots-l_n\zeta_n}\,
\vev{\tilde{\Psi}(l_1)\cdots\tilde{\Psi}(l_n)} \n\\
&=&\frac{
\left\langle\left.\left.\left.\,
-\frac{B}{g}\,\right|\,
:\dphi_0(\zeta_1)\cdots\dphi_0(\zeta_n):
\,\right|\,\Phi\,\right\rangle\right.}{
\left\langle\left.\left.\,
-\frac{B}{g}\,\right|\,
\,\Phi\,\right\rangle\right.},
\eeqa
and the connected correlation functions are obtained
as their cumulants.
Here the normal ordering respects $SL(2,\bC)$-invariant vacuum,
and the state $\bra{-B/g} = \bra{\hat{\sigma}}
\exp\left\{-(1/g)\sum_{n=1}^{2p+1}B_n \alpha_n\right\}$
characterizes the theory
and corresponds schematically to the NL$\sigma$-model 
action $S = \sum_{n=1}^{2p+1}B_n \cO_n$.
The state $\pvac$ satisfies the $W_{1+\infty}$ constraints
\cite{fkn1-dvv,gn,g,fkn2-fkn3}:
\beqa
W^k_n\,\pvac=0~~~(k\geq1,~n\geq -k+1).
\eeqa
The generators of the $W_{1+\infty}$ algebra \cite{winf}
are given by the mode expansion
\beqa
W^k(\zeta)\,\equiv\,\sum_{n\in\bZ}W^k_n\,\zeta^{-n-k}
\,=\, \sum_{a=0}^{p-1}:\cb_a(\zeta)\,\del_\zeta^{k-1}\,c_a(\zeta):.
\eeqa
Here $c_a(\zeta)$ and $\cb_a(\zeta)$ $(a=0,1,...,p-1)$
are fermions constructed from the scalars by bosonization:
\beqa
\cb_a(\zeta)\,=\,K_a\,:e^{\rphi_a(\zeta)}:,~~~~~~~
c_a(\zeta)\,=\,K_a\,:e^{-\rphi_a(\zeta)}:,
\eeqa
where $K_a$ is a cocycle factor that ensures the correct
anticommutation
relations between different indices $a\neq b$,
and all the operators are again normal-ordered with respect to
$SL(2,\bC)$
invariant vacuum.
In addition to the $W_{1+\infty}$ constraints,
we further require that $\pvac$ be a decomposable state.\footnote{
A state $\pvac$ is called decomposable
if it is written as $\pvac=e^H \ket{\hat{\sigma}}$,
where $H$ is a bilinear form of the fermions.
This is equivalent to the statement that
$\tau(x)=\bra{\hat{\sigma}} \exp\{\sum_{n=1}^\infty x_n\alpha_n\}
\pvac$ is a $\tau$ function of the KP hierarchy \cite{djkm}.
It is proved in \cite{fkn2-fkn3} that
this set of conditions ($W_{1+\infty}$ constraints and
decomposability) is equivalent to
the Douglas equation \cite{md}, $[P,\,Q]=1$.
}

Now we try to construct the D-instanton operator
in this operator formalism.
By definition it should be expressed as an operator
that specifies a point $\zeta$ at which
boundaries of world-sheets are glued together (see Fig.\ 1).
\begin{figure}
\begin{center}
\leavevmode
\epsfbox{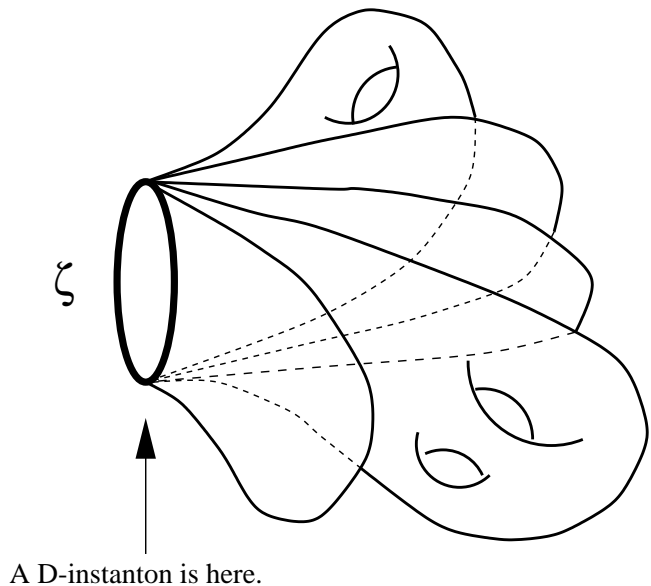}
\end{center}
\caption{Geometrical meaning of the D-instanton operator.
All the points on the boundary are mapped to 
a single point $\zeta$ in the target space.}
\end{figure}
Since there exists a redundancy reflecting the invariance
under one-dimensional diffeomorphism along each boundary,
this should be gauge-fixed if one would like to count
only independent configurations.
This can be simply done by creating strings at $\zeta$
not with $\dphi_0(\zeta)$ but with $-\rphi_0(\zeta)$,
since for the latter the redundancy that is proportional to
loop length is automatically divided out:
$-\rphi_0(\zeta)\sim \int_0^\infty dl
e^{-l\zeta}\,(1/l)\,\tilde{\Psi}(l)$.
Taking also into account the monodromy of scalar fields,
we should equally treat all the $-\rphi_a(\zeta)$'s
as such gauge-fixed string fields.
Thus, after summing over the number of the boundaries glued at
$\zeta$,
we have $\sum_{n}(q_{a}^n/n!)\left(-\rphi_a(\zeta)\right)^n
\,=\,\exp\left\{-q_a\rphi_a(\zeta)\right\}$,
where $1/n!$ is a statistical factor
and we assume that a boundary of a connected world-sheet is
accompanied with weight $q_a$ for each $-\rphi_a(\zeta)$.
{}Furthermore, we need to make an integration over the collective 
coordinate $\zeta$ of D-instanton. 
The path of the integration surrounds $\zeta=\infty$ $p$ times,
which corresponds to the integration over $X$ 
with $0\leq X\leq 2\pi/\alpha_0$. 
We are thus lead to the following partition function of
$N$ D-instantons:
\beqa
Z_N~=~\gint d\zeta_1 \cdots \gint d\zeta_N
\vev{\prod_{i=1}^{N} e^{- \vec{q}\cdot 
\vec{\rphi}(\zeta_i)}} ,\label{10}
\eeqa
where $\vec{q}\cdot \vec{\rphi}(\zeta_i)=\sum_{a=0}^{p-1}
q_a \rphi_a(\zeta_i)$ and
$\zeta_i$ represents the collective coordinate of the
$i$-th D-instanton.
Note that this equation actually realizes
Polchinski's combinatorics of boundaries \cite{p-g}.
After summing over $N$, we obtain the grand-canonical partition 
function:
\beqa
Z&=& \sum_{N=0}^\infty\,\frac{\theta^N}{N!}\,Z_N \nn
&=&\vev{ \exp\left\{\theta \gint d\zeta\,\,
e^{-\vec{q}\cdot \vec{\rphi}(\zeta)}\right\}}, \label{11}
\eeqa
where $1/N!$ is a statistical factor and $\theta$ is a fugacity.
Equations (\ref{10}), (\ref{11}) imply that our D-instanton can be 
identified with an object which locally couples to the scalar fields
$\rphi_a(\zeta)$ at the position $\zeta$ with ``charges'' {$q_a$}.

This is not the end of story. 
In fact, the charges {$q_a$} in (\ref{11}) 
cannot take arbitrary values, 
since in order for the expression to give a correct background,
the state $\exp\left\{\theta \gint d\zeta\,
e^{-\vec{q}\cdot \vec{\rphi}(\zeta)}\right\}\pvac$
must satisfy the $W_{1+\infty}$ constraints when $\pvac$ does.
This is equivalent to the condition that the operator
$\gint d\zeta\,e^{-\vec{q}\cdot \vec{\rphi}(\zeta)}$
should commute with any generators of the $W_{1+\infty}$ algebra,
and it is shown in \cite{fy1} that, for this condition to hold,
only two charges among {$q_a$} can take nonvanishing values
with $\pm 1$.
This is actually the combination for which the operator 
can be expressed as a fermion bilinear, 
and has exactly the same form with the operator $D_{ab}$ 
that was introduced in \cite{fy1,fy2}:
\beqa
D_{ab}&=& \gint\ddz\,\cb_a(\zeta)c_b(\zeta)~\sim~
\gint\ddz\,e^{\rphi_a(\zeta)-\rphi_b(\zeta)} ~~~~(a\neq b).
\eeqa
{}Furthermore, since there are various ways to choose the pairs 
$a$ and $b$,
we can introduce the corresponding fugacities $\theta_{ab}$, 
and finally have the following form instead of
\eq{11}:
\beqa
Z~=~\vev{\prod_{a\neq b} \exp\left\{\theta_{ab}D_{ab}\right\}}
~=~\vev{\prod_{a\neq b} \exp\left\{\theta_{ab} \gint d\zeta\,
e^{\rphi_a(\zeta)-\rphi_b(\zeta)}\right\}} .
\eeqa
Thus, we conclude that the $D_{ab}$ is the operator that creates
a D-instanton and the $\prod_{a\neq b}e^{\theta_{ab}D_{ab}}$
the creation operator of multi D-instantons.
Notice that in \cite{fy2} the form of the multi D-instanton operator
was originally determined by requiring that the decomposability
be preserved when acting on $\pvac$.

In the rest of this article, 
we briefly explain how the nonperturbative effects due to 
D-instantons can be calculated in our operator formalism, 
and show that they coincide with those effects that were 
found in exact solutions of string equations. 
Most of the argument here will closely follow \cite{fy1,fy2}, 
and to make the comparison most easily, 
we mainly consider the string susceptibility 
in the D-instanton background: 
\beqa
u(t,g,\,\theta)&\equiv& g^2\,\del_t^{\,2}\,\log\,Z\nn
&=&g^2\,\del_t^{\,2}\,
\log\,\left.\left\langle\left.-\frac{B}{g}\,\right|\,
\prod_{a\neq b}e^{\,\theta_{ab}\,D_{ab}}
\,\right|\,\Phi\,\right\rangle, 
\eeqa
where $t$ and $g$ are the cosmological and string coupling 
constant, respectively, 
and we have set $B_1=t,~B_{2p+1}=-4p/((p+1)(2p+1)),~
B_n=0~(n\neq 1,~2p+1)$. 
The free energy $\log \left\langle\left.-B/g\,\left|\,
\prod_{a\neq b}e^{\,\theta_{ab}\,D_{ab}}\,\right|\right.\,\Phi\,
\right\rangle$ in (14) can be evaluated around $\theta=0$ 
by rewriting it as follows:
\beqa
\log\,\left.\left\langle\left.-\frac{B}{g}\,\right|\,
\prod_{a\neq b}e^{\,\theta_{ab}\,D_{ab}}
\,\right|\,\Phi\,\right\rangle
\,=\,\log\left\langle\left.-\frac{B}{g}\,\right|\,\Phi\,\right\rangle
\,+\,\log\,\vev{\prod_{a\neq b}e^{\,\theta_{ab}\,D_{ab}}},
\eeqa
where
\beqa
\vev{\prod_{a\neq b}e^{\,\theta_{ab}\,D_{ab}}}\,=\,
\frac{
\left\langle\left.\left.\left.\,
-\frac{B}{g}\,\right|\,
\prod_{a\neq b}e^{\,\theta_{ab}\,D_{ab}}
\,\right|\,\Phi\,\right\rangle\right.}{
\left\langle\left.\left.\,
-\frac{B}{g}\,\right|\,
\,\Phi\,\right\rangle\right.}.
\eeqa
Thus, expanding $e^{\theta_{ab} D_{ab}}$ in (15) with respect to
$\theta_{ab}$ and picking up only the first-order term, 
we get the string susceptibility which includes one-D-instanton effect:
\beqa
u(t,g,\theta)\,=\,u_{\rm pert}(t,g)
+ g^2 \sum_{a\neq b} \theta_{ab}
\del_t^{\,2} \vev{D_{ab}}
+O(\theta_{ab}^2).
\eeqa
Here $u_{\rm pert}(t,g)$ is the string susceptibility 
which is perturbatively evaluated at $\theta_{ab}=0$.
We will make a comment on multi-D-instanton effects later.

The expectation value of $D_{ab}$ in (17) can be represented as
\beqa
\vev{D_{ab}}&=&\gint\ddz \vev{e^{\rphi_a(\zeta)-\rphi_b
(\zeta)}}\n\\
&=&\gint\ddz \exp\left\{\vev{e^{\rphi_a(\zeta)-\rphi_b
(\zeta)}-1}_c
\right\}\n\\
&=&\gint\ddz \exp\left\{
\vev{\rphi_a(\zeta)-\rphi_b(\zeta)}
+\frac{1}{2}\vev{\left(\rphi_a(\zeta)-\rphi_b
(\zeta)\right)^2}_c
+\,\cdots\right\}.
\eeqa
Since a connected $n$-point function has the following expansion in
$g$: 
\beqa
\vev{\rphi_{a_1}(\zeta_1)\cdots\rphi_{a_n}(\zeta_n)}_c
=\sum_{h=0}^\infty 
\vev{\rphi_{a_1}(\zeta_1)\cdots\rphi_{a_n}(\zeta_n)}_c^{(h)}
g^{-2+2h+n},
\eeqa
we know that in the weak coupling limit, 
leading contributions to the exponent
come from spherical topology ($h=0$):
\beqa
  \vev{D_{ab}}&=&\gint\ddz\,
    e^{\,(1/g)\,\Gamma_{ab}(\zeta)
    \,+\,(1/2)\,K_{ab}(\zeta) \,+\,O(g)}
\eeqa
with $\Gamma_{ab}(\zeta)\,\equiv\,\vev{\rphi_a(\zeta)\,-\,
\rphi_b(\zeta)}^{(0)}$ and 
$K_{ab}(\zeta)\,\equiv\,\vev{(\rphi_a(\zeta)\,-\,
\rphi_b(\zeta))^2}_c^{(0)}$.
These functions $\Gamma_{ab}(\zeta)$ and $K_{ab}(\zeta)$ 
can be calculated by integrating the disk and cylinder amplitudes 
($\vev{\Psi(\zeta)}^{(0)}$ and 
$\vev{\Psi(\zeta_1)\Psi(\zeta_2)}_c^{(0)}$) 
\cite{bdss} followed by analytic continuation. 
{}For example, $\Gamma_{ab}(\zeta)$ can be evaluated as 
\beqa
\Gamma_{ab}(\zeta)&=&
\vev{\rphi_a(\zeta)}^{(0)}-\vev{\rphi_b(\zeta)}^{(0)}\nn
&=&\vev{\rphi_0({e}^{2\pi ia}\zeta)}^{(0)}
-\vev{\rphi_0({e}^{2\pi ib}\zeta)}^{(0)}\nn
&=&\left.\int^\zeta
d\zeta'\vev{\Psi(\zeta')}^{(0)}
\right|_{\zeta\rightarrow {e}^{2\pi ia}\zeta}
\,-\, \left.\int^\zeta 
d\zeta'\vev{\Psi(\zeta')}^{(0)}
\right|_{\zeta\rightarrow {e}^{2\pi ib}\zeta} .
\eeqa
Thus, the leading contribution in the weak coupling limit 
can be calculated by applying the saddle point method 
in the complex $\zeta$ plane. 
To do so, it is convenient to introduce a new coordinate $s$ which is
defined by
\beqa
s(\zeta)\,=\,\frac{1}{\sqrt{t}}\left(\zeta+\sqrt{\zeta^2-t}\right),
\eeqa
for which $\Gamma_{ab}(s)$ is expressed as 
\beqa
  \Gamma_{ab}(s)&=&\gamma_p\,\left[\, \frac{1}{r+1}\left\{
  (\omega^a-\omega^b)s^{r+1}+
  (\omega^{-a}-\omega^{-b})s^{-(r+1)}\right\}
  \right.\n\\
  &~&~~~~~-\,\left.\frac{1}{r-1}\left\{
  (\omega^a-\omega^b)s^{r-1}+
  (\omega^{-a}-\omega^{-b})s^{-(r-1)}\right\}
  \,\right],
\eeqa
where $\gamma_p=2^{-1/p}\,t^{(2p+1)/2p} /(p+1)$,
$\omega=e^{2\pi i/p}$ and $r=(p+1)/p$.
The saddle points are found at $s_0\,=\,e^{(np-a-b)\pi i/(p+1)}$, 
which gives 
\beqa
\Gamma_{ab}(s_0)&=&\frac{8p}{2^{1/p}\cdot (2p+1)}
\,t^{\frac{2p+1}{2p}}
\sin\left(\frac{a+b+n}{p+1}\pi\right)\,
\sin\left(\frac{a-b}{p}\pi\right).\label{24}
\eeqa
Here $n$ takes integers which satisfy $\Gamma_{ab}(s_0)<0$ 
(see \cite{fy1, fy2} for detailed investigations).
They give the leading nonperturbative effects that are 
proportional to $e^{(1/g)\,\Gamma_{ab}(s_0)}$. 
Coefficients can also be calculated explicitly by performing 
an integration around the saddle points (see \cite{fy1, fy2}).

Next, we compare these nonperturbative effects 
with those that were found in the exact solutions 
of string equations.
In the $p=2$ (pure gravity) case, the string equation is
$4u^2 + (2g^2/3)\del_t^2u = t$ (Painlev\'{e} I equation) 
\cite{bk-ds-gm}.
The leading nonperturbative effect can be calculated 
by expanding the equation around $u_{\rm pert}=-\sqrt{t}/2+O(g^2)$ 
and is found to be $\Delta u\equiv u-u_{\rm pert}\,\propto\,
e^{C/g}$ with $C=-4\sqrt{6}\,t^{5/4}/5$ \cite{s, bk-ds-gm, d}.
This exponent coincides with $\Gamma_{01}(s_0)$ with $n=0,\,1$ 
and $\Gamma_{10}(s_0)$ with $n=3,\,4$ in (\ref{24}).

{}For the $p=3$ (Ising) case, the string equation is
$4 u^3 + (3g^2/2)(\del_tu)^2
+ 3g^2 u \del_t^2u + (g^4/6) \del_t^4u = - t$ 
\cite{ising}.
The leading nonperturbative effects are of the form 
$e^{C/g}$ with $C=-6\sqrt{6}\,t^{7/6}/
(2^{1/3}\cdot7)$ or
$-12\sqrt{3}\,t^{7/6}/(2^{1/3}\cdot7)$. 
The set of these exponents agrees
with that of different negative values of $\Gamma_{ab}(s_0)$
in (\ref{24}).

We finally consider the $p=4$ (tricritical Ising) case. 
The string equations are now coupled differential equations
for two unknown functions $u$ and $v$ \cite{tcising}:
\beqa
  0&=&40u^3+40uv+15g^2(\del_tu)^2+10g^2u{\del_t}^2u
    +10g^2{\del_t}^2v-2g^4{\del_t}^4u \\
  5t&=&-20u^4+40v^2+50g^2u(\del_tu)^2+20g^2\del_tu\del_tv
    +20g^2u^2{\del_t}^2u \n\\
  &&~~~~ -20g^2u{\del_t}^2v
    +11g^4({\del_t}^2u)^2+13g^4\del_tu{\del_t}^3u
    +11g^4u{\del_t}^4u+g^4{\del_t}^4v.
\eeqa
The leading nonperturbative effects are of the form $e^{C/g}$
with $C=-2^{13/4}\sqrt{5\pm \sqrt{5}}\,t^{9/8}/9$ or
$-2^{11/4}\sqrt{5\pm \sqrt{5}}\,t^{9/8}/9$.
The set of these exponents completely accord with that of 
different negative values of $\Gamma_{ab}(s_0)$ in (\ref{24}).

Moreover, we can also evaluate multi-D-instanton effects
by using our formalism. As an example, the $p=2$ (pure gravity)
case was considered in detail in \cite{fy2} 
and the string susceptibility in the multi-D-instanton background 
was obtained as
\beqa
u(t,g,\,\theta)\,=\,-\,\frac{\sqrt{t}}{2}\,+\,
6\,\theta_{\rm R}\,\sqrt{g}\,t^{-1/8}\,
e^{-\,4\sqrt{6}\,t^{5/4}/\,5g}\,\times \,
\left(1\,+\,\theta_{\rm R}\,\sqrt{g}\,t^{-5/8}
e^{-\,4\sqrt{6}\,t^{5/4}/\,5g}\right)^{-2}.
\eeqa
Here we have neglected contributions from higher topologies,
and $\theta_{\rm R}$ is the renormalized fugacity 
which absorbed an integration constant that arises 
when integrating cylinder amplitudes.
This $u(t,g,\,\theta)$ exactly reproduces 
a series of nonperturbative corrections in the string equation 
for pure gravity \cite{fy2}.

These examples confirm that the stringy nonperturbative effects 
found in the exact solutions of string equations 
can be interpreted as D-instanton effects.

In this short letter, we demonstrate that the conformal field theory
that used to be a technical tool to compactly describe
macroscopic-loop amplitudes, is actually the field theory
that describes the target space
where the boundaries of world-sheets live.
We also explicitly construct 
analogues of D-instantons which satisfy 
Polchinski's ``combinatorics of boundaries.''
It is surprising that our operator formalism can
be naturally applied to this combinatorics.
Moreover, we show that this formalism is
powerful in evaluating the D-instanton effects
and demonstrate for $p=2, 3, 4$ that these effects
coincide with the stringy nonperturbative effects found
in the exact solutions of string equations.
{}Finally we point out the similarities between noncritical strings
and the Sine-Gordon theory
(or more generally affine $SU(p)$ Toda field theories)
since for both the fundamental degrees of freedom
are described by scalar fields $\rphi_a$
corresponding to the fundamental weights
(this is actually $SU(p)$, not $U(p)$, because 
$\sum_{a=0}^{p-1}\dphi_a= 0$ under the $W_{1+\infty}$ constraints),
while the solitons are expressed by their exponentials
in a combination associated with the roots,
$\exp\left(\rphi_a-\rphi_b\right)$.

%%%%%%%%%%%%%%%%%%%%%%%%%%%%%%%%%%%%%%%%%%%%%
% Acknowledgment
%%%%%%%%%%%%%%%%%%%%%%%%%%%%%%%%%%%%%%%%%%%%
\section*{\protect\large{\protect\bf Acknowledgment}}
The authors would like to thank K.\ Hamada, S.\ Hirano, H.\ Itoyama,
V.\ Kazakov, I.\ Kostov, Y.-S.\ Wu, T.\ Wynter and T.\ Yoneya
for useful discussions.
One of the authors (M.F.) also would like to thank Saclay
and the lab of theoretical physics of Ecole Normale 
Sup\'{e}rieure 
for their hospitality.
This work is supported in part by the Grant-in-Aid for Scientific
Research from the Ministry of Education, Science and Culture,
and by the Sumitomo Foundation.

\newpage
%%%%%%%%%%%%%%%%%%%%%%%%%%%%%%%%%%%%%%%%%%%%%%
% References
%%%%%%%%%%%%%%%%%%%%%%%%%%%%%%%%%%%%%%%%%%%%%%

\end{document}